\documentclass[10pt]{iopart}
\input epsf
\usepackage{graphicx}

\begin{document}

\title{Frustrated magnets in three dimensions: a nonperturbative approach}

\author{B. Delamotte\dag, D. Mouhanna\dag\ and M. Tissier\ddag}  
\address{\dag\ Laboratoire de Physique Th\'eorique et Hautes Energies 
 
Universit\'e Paris VI-Pierre et Marie Curie, Universit\'e  Paris VII-Denis Diderot and UMR  7589 of CNRS, 2 Place Jussieu, 75251 Paris Cedex 05, France}

\address{\ddag\ Laboratoire de Physique Th\'eorique des Liquides 

Universit\'e Paris VI-Pierre et Marie Curie and UMR  7600  of CNRS, 4 Place Jussieu, 75252 Paris
 Cedex 05, France}

\begin{abstract}

Frustrated  magnets exhibit unusual  critical behaviors:  they display
scaling  laws accompanied  by {\it  nonuniversal}  critical exponents.
This suggests  that these systems {\it generically}  undergo very weak
first  order phase transitions.  Moreover, the  different perturbative
approaches  used to  investigate  them  are in  conflict  and fail  to
correctly reproduce their  behavior.  Using a nonperturbative approach
we explain the mismatch  between the different perturbative approaches
and account for the {\it nonuniversal scaling} observed.

\end{abstract}


\section{Introduction}

After twenty  five years  of investigations, the  nature of  the phase
transition undergone  by frustrated magnets is  still strongly debated
\cite{diep94,kawamura98}.  The main feature  of these systems is that,
due to  frustration, their ground state displays  a {\it noncollinear}
order.  This  is  the  case  for  the  celebrated  Stacked  Triangular
Antiferromagnets  (STA) whose  ground state  exhibits  a 120$^{\circ}$
structure.   As a  consequence, for  $N=3$ for  instance,  the $SO(3)$
symmetry group of the high-temperature  phase is {\it fully} broken in
the low-temperature one.  Also, the order parameter is no longer given
by a  vector, as in  the nonfrustrated case,  but by a matrix  made of
{\it  two}  orthonormal  vectors  $\vec{\phi}_1$  and  $\vec{\phi}_2$.
These characteristics have  led to a conjecture of the existence  of a {\it
new}  universality   class  for  frustrated  magnets.   In  fact,  the
interpretation as  well as the  description of their  critical physics
appear  to  be  considerably  more  involved  than  for  nonfrustrated
magnets:  first,  because  there   are  several   experimental  facts
incompatible with  a universal critical behavior;  second, because the
different theoretical  approaches used to analyze  their behaviors are
in conflict and fail to reproduce the experimental data.

\section{Experimental and numerical contexts}

The first indications of a nontrivial critical behavior for frustrated
magnets  appear in  the  analysis of  the  experimental and  numerical
data.  An extensive  analysis  of  the compounds  and  systems  supposed 
{\it  a priori}  to belong  to the same  universality class  as  the
STA, such as   Helimagnets, has been  performed by the present  authors in
\cite{tissier01,delamotte03}.  Without entering  into  the details  we
give the  main features that  have emerged from this  analysis. First,
scaling laws are found in most compounds. This is also the case within
all  ---   but  an   important  one  \cite{itakura03}   ---  numerical
simulations  of the  STA.  This  scaling behavior  is  associated with
critical   exponents  that   differ   from  those   of  the   standard
$O(N)/O(N-1)$ universality  classes. This fact has been  at the origin
of  the  belief  that   frustrated  magnets  could  constitute  a  new
universality class. However, there  are other facts characterizing the
critical behavior  of both XY  and Heisenberg systems that  go against
this belief. 

{\it  i)} There are two groups  of compounds that clearly
differ by their critical exponents. 

 {\it ii)} The anomalous dimension
$\eta$  is {\it  negative} for  many  compounds and  for many  systems
numerically simulated.  We recall  that the anomalous dimension $\eta$
is {\it  always} positive if the  phase transition is  truly of second
order and  if the underlying field theory  is a Ginzburg-Landau-Wilson
(GLW) $\phi^4$-like  theory, as  it is the  case here.

 {\it  iii)} The
scaling relations are violated by several standard deviations for many
compounds.

 {\it iv)} Models  differing from STA by their microscopical
details undergo either strong  first order phase transitions (XY case)
\cite{itakura03,loison98}  or exhibit  scaling  behaviors incompatible
with   that   of   the   numerical   STA   model   (Heisenberg   case)
\cite{loison00b,kunz93,loison99}.  

 {\it v)}  Recent  Monte Carlo  and
Monte Carlo RG approaches of STA and GLW models give clear indications
of first order behaviors \cite{itakura03}.

From all  these observations,  it appears that  an explanation  of the
critical  physics of  frustrated  magnets assuming  a standard  second
order  critical  behavior  relying   on  the  existence  of  a  unique
universality class is very likely excluded.

\section{Field theoretical context}

\subsection{The nonlinear sigma model approach}

A first  theoretical approach consists in  a low-temperature expansion
performed  near $d=2$ on  the suitable  field theory,  the $O(3)\times
O(2)/         O(2)$        nonlinear         sigma        (NL$\sigma$)
model~\cite{azaria90,azaria93}  which is  obtained  by expressing  the
interaction  between  spins  in  terms  of the  Goldstone  modes.  The
renormalization  of the  interaction  between these  modes is  treated
within   a  double   expansion   in  the   temperature   $T$  and   in
$\epsilon=d-2$. The result  of this study is that  there exists, above
$d=2$,  a stable fixed  point~\cite{azaria90,azaria93}. At  this fixed
point,  the  symmetry breaking  scheme  is  enlarged from  $O(3)\times
O(2)/O(2)$ to  $O(3)\times O(3)/O(3)$ which  itself has the  same {\it
local} structure as $O(4)/O(3)$ on which is built the NL$\sigma$ model
relevant to four-component {\it nonfrustrated} spins. As seen from the
NL$\sigma$  model  approach  ---  that  only  depends  on  this  local
structure  ---   the  two  above   theories  are  equivalent   to  all
orders. Accordingly, the critical  properties of the STA model between
$d=2$  and  $d=4$  should  be  governed by  the  standard  $O(4)/O(3)$
universality  class~\cite{azaria90,azaria93}.  This  is,  however,  in
disagreement with both the experimental  and numerical data as well as
the  weak-coupling analysis  of the  GLW model  of  frustrated magnets
around  $d=4$. Therefore, there  must exist  nonperturbative phenomena
--- with respect  to the  expansion parameters $T$  and $\epsilon=d-2$
--- leading  to  the  breakdown   of  the  NL$\sigma$  model  approach
somewhere between $d=2$ and $d=3$.

\subsection{The GLW model approach}

An alternative approach is  to consider a Ginzburg-Landau-Wilson (GLW)
 representation    of    the    physics    of    frustrated    magnets
 \cite{jones76,garel76,bailin77,yosefin85,kawamura88}.   This  can  be
 obtained by  relaxing, via a potential, the  constraints defining the
 ground state. The resulting Hamiltonian writes:
\begin{equation}
\begin{array}{ll}
H=\displaystyle  \int      d^dx       &  \displaystyle    \bigg\{
{1\over2}\left((\partial      {\vec{\phi_1})}^2  \right.   \left. +      {(\partial
\vec{\phi_2})}^2\right)                   +                  {m^2\over
2}\left(\vec{\phi_1}^2+\vec{\phi_2}^2\right)    +\\     &     +     u
\left(\vec{\phi_1}^2+\vec{\phi_2}^2\right)^2                          +
v\bigg(\vec{\phi_1}^2\vec{\phi_2}^2-{(\vec{\phi_1}.\vec{\phi_2})}^2\bigg)\bigg\}\
.
\label{GLW}
\end{array}
\end{equation}
The  RG equations for  the coupling  constants $u$  and $v$  have been
first   determined   in  a   double   expansion   in   $u$,  $v$   and
$\epsilon=4-d$.  At  leading  order~\cite{garel76,bailin77,yosefin85},
they  display {\it no}  stable fixed  point in  the XY  and Heisenberg
cases. A  first order phase transition  is thus expected  in $d=3$, in
contradiction with the NL$\sigma$ model results and with the existence
of   scaling   behaviors   for   frustrated  magnets.   An   important
generalization   of  the   GLW  model   (\ref{GLW})  is   to  consider
$\vec{\phi}_1$    and   $\vec{\phi}_2$   as    $N$-component   vectors
~\cite{garel76,bailin77,yosefin85}.  For $N$  larger  than a  critical
value $N_c(d)$  depending on the  dimension, the RG  equations display
two nontrivial  fixed points:  one, $C_+$, is  stable; the  other one,
$C_-$, is  unstable. Above $N_c(d)$, the transition  is thus predicted
to  be of  second order.  As $N$  is lowered  starting from  values of
$N>N_c(d)$,  the two  fixed  points  $C_+$ and  $C_-$  get closer  and
finally collapse together for  $N=N_c(d)$. Below $N_c(d)$, there is no
longer a  stable fixed point and the  transition is expected to be of first
order.   The   value  of  $N_c(d)$   has  been  computed   by  several
perturbative   means:  in  a   double  expansion   in  $u$,   $v$  and
$\epsilon=4-d$  up to five-loops  \cite{calabrese03c} and  directly in
$d=3$    in    a    weak-coupling    expansion   up    to    six-loops
\cite{pelissetto01a}.  These  computations lead  to  an almost  common
value  $N_c(d=3)\simeq 6$.  This result  could be  interpreted  as the
existence of a  first order phase transition in $d=3$  for both XY and
Heisenberg  spins.    This  is  only  true   within  the  $4-\epsilon$
expansion.  In  effect ---  and  rather  surprisingly  --- within  the
six-loop weak-coupling expansion performed in $d=3$ by Pelissetto {\it
et  al} \cite{pelissetto01a}, a fixed  point is  found for  $N=2$ and  $N=3$.
 It  has been
conjectured by Calabrese  {\it et al} \cite{calabrese02,calabrese03b}
that  these  fixed points  could  explain  the  spreading of  critical
exponents  encountered  in  frustrated  magnets. Indeed,  due  to  the
specific structure of the RG flow around the {\it focus} fixed
point, the  critical exponents display strong variations  along the RG
trajectories  that could  explain  the lack  of universality  observed
experimentally  and   numerically.   Let  us   now  underline  several
drawbacks  of the scenario  of Calabrese  {\it et  al}. First,  it is
based on  the existence of  fixed points that  are not related  to any
already known fixed points. In  particular, the fixed points found for
$N=2$ and $N=3$ within this computation in $d=3$ that would govern the
critical physics  of frustrated  magnets are, according  to Pelissetto
{\it  et al}   and Calabrese  {\it  et al},  {\it nonanalytically}
related to those found in the large-$N$ as well as in the $4-\epsilon$
expansions. This means  that there is no way  to check their existence
using  these perturbative  methods. Second,  it is  very  difficult to
account, in this framework, for  the first order behavior deduced from
several numerical simulations of XY and Heisenberg systems.  Third, it
is  also very difficult  to explain  why there  is no  physical system
characterized by  the exponents associated with the  fixed point found
by Pelissetto {\it  et al}.  Finally, there is  no explanation of the
breakdown of the NL$\sigma$ model predictions.

\section{Effective average action approach}

To clarify this situation  we have employed a nonperturbative approach
based    on    the     concept    of    effective    average    action
\cite{wetterich91,berges02},  $\Gamma_k[\phi]$,   which  is  a  coarse
grained free energy  in which only fluctuations with  momenta $q\ge k$
have been integrated out.  The field $\phi$ stands here for an average
order parameter at  the running scale $k$. In the  limit $k\to 0$, all
modes are  taken into account  and $\Gamma_{k=0}$ identifies  with the
usual Gibbs free energy, {\it  i.e.} the effective action $\Gamma$. At
the scale $k=\Lambda$  --- the overall cut-off ---  no fluctuation has
been   integrated  and   $\Gamma_{k=\Lambda}$   identifies  with   the
microscopic   Hamiltonian.  The   $k$-dependence   of  $\Gamma_k$   is
controlled       by       an       exact      evolution       equation
\cite{wetterich93c,tetradis94}:
\begin{equation}
{\partial    \Gamma_k\over    \partial    t}={1\over   2}    \hbox{Tr}
\left\{{\partial   R_k\over  \partial   t}{1\over  \Gamma_k^{(2)}+R_k}
\right\}\ ,
\label{renorm}
\end{equation}
where  $t=\ln  \displaystyle {k  /  \Lambda}$.  The  trace has  to  be
 understood  as  a momentum  integral  as  well  as a  summation  over
 internal indices. In equation (\ref{renorm}), $\Gamma_k^{(2)}$ is the
 {\it exact  field-dependent} inverse  propagator --- i.e.  the second
 derivative  of $\Gamma_k$  --- and  $R_k$ is  the  effective infrared
 cut-off  which  suppresses  the  propagation of  modes  with  momenta
 $q<k$.  The  effective average  action  $\Gamma_k$  is a  functional,
 invariant under  the symmetry group  of the system and  thus includes
 all powers of all invariants --- and their derivatives --- built from
 the  average  order parameter.  Thus,  equation  (\ref{renorm}) is  a
 nonlinear functional equation, too  difficult to solve exactly so
 that $\Gamma_k$  must be truncated. One ---  fruitful --- possibility
 is to perform a double  expansion of $\Gamma_k$ in derivatives and in
 fields  keeping a  finite number  of  monomials in  the fields  while
 including the most relevant derivative terms~\cite{berges02}. We have
 first              considered              the              following
 truncation~{\cite{tissier01,delamotte03,tissier00}:
\begin{equation}
\hspace{-1cm}
\begin{array}{ll}
\Gamma_k[\vec\phi_1,\vec\phi_2]=&      \displaystyle     \int     d^dx
\bigg\{\frac{{Z_k}}{2}\left(\left(\partial\vec
\phi_1\right)^2+\left(\partial   \vec  \phi_2\right)^2\right)   +\\  &
\displaystyle    \frac{\omega_k}{4}   \left(\vec   \phi_1\cdot\partial
\vec\phi_2-\vec         \phi_2\cdot\partial        \vec\phi_1\right)^2
+\frac{{\lambda_k}}{4}\left(\frac\rho     2    -\kappa_k\right)^2    +
\frac{{\mu_k}}{4} \tau \ \bigg\}
\label{action}
\end{array}
\end{equation}
where  $\left\{\omega_k, \lambda_k,  \kappa_k,  \mu_k,Z_k\right\}$ are
the  coupling constants that  parametrize the  evolution of  the model
with   the   scale,   while   $\rho={\hbox{Tr}}\   ^{t}\phi\phi$   and
$\tau={1\over  2}{\hbox{Tr}}  (^{t}\phi\phi)^2-{1\over 4}({\hbox{Tr}}\
^{t}\phi\phi)^2$ are the  two independent $O(N)\times O(2)$ invariants
built  from  the  average  order parameter  $\phi=\phi_{ab}$,  a  real
$N\times  2$ matrix  that  gathers the  two  vectors $\vec\phi_1$  and
$\vec\phi_2$.  This ansatz  realizes the  symmetry breaking  scheme of
frustrated magnets: $O(N)\times  O(2)\to O(N-2)\times O(2)_{diag}$. We
have  derived the  nonperturbative  flow equations  for the  different
coupling     constants    using    equations     (\ref{renorm})    and
(\ref{action}). We  have also checked  the convergence of  our results
with respect  to the order of  the truncation of  $\Gamma_k$ in $\rho$
and  $\tau$.  The  explicit   recursion  equations  for  the  coupling
constants     are    too    lengthy     to    be     displayed    here
{\cite{tissier01,delamotte03,tissier00}  so   we  concentrate  on  the
physical results.

\subsection{Fixed point structure}

 We  have  numerically analyzed  our  flow  equations  by varying  the
dimension  between  $d=2$  and  $d=4$  for different  values  of  $N$,
identified  the different  fixed points  and studied  their stability.
Around $d=2$ we find, for all  $N>2$, a stable fixed point $C_+$ which
identifies  with the  NL$\sigma$ model  fixed point.  For sufficiently
large values of  $N$ --- $N>21.8$ --- one can  follow this fixed point
up to $d=4$ where the  one-loop results of the weak-coupling expansion
performed around  this dimension  are recovered. We  are thus  able to
smoothly  interpolate  between the  perturbative  results obtained  in
$d=2$ and  $d=4$. On the  other hand, for any  value  of $N>2$,
increasing $d$, one finds a {\it new and unstable} fixed point $C_-$ in
a  dimension $d_1(N)$.  For any value of $N<21.8$ and as $d$  is  further
increased,  the two  fixed
points $C_+$ and $C_-$ get closer, coalesce and finally become complex
in a dimension $d_2(N)$.  Above this dimension $d_2(N)$, no real fixed
point is  found, a fact  which is interpreted  as the sign of  a first
order behavior. This collapse of the fixed points $C_+$ and $C_-$, for
different  values of  $N$, generates  the curve  $N_c(d)$.  As already
said, for large values of $N$  --- typically $N$ of order 21.8 --- the
collapse occurs  around $d=4$ in  agreement with the ---  standard ---
weak-coupling analysis performed  around this dimension.  However, for
low values of  $N$ --- $N=3$ for instance ---,  this collapse of fixed
points  operates  between $d=2$  and  $d=3$:  one has  $d_2(N=3)\simeq
2.8$. This means that our approach provides a nonperturbative solution
to    the    breakdown    of    the    NL$\sigma$    model    approach
{\cite{tissier01,tissier00,tissier00b}:  the stable  $O(4)/O(3)$ fixed
point obtained  within this approach  collapses in $d\simeq  2.8$ with
another ---  unstable ---  fixed point  which is {\it  not} seen  in a
low-temperature expansion.

\subsection{The physics in $d=3$}

We now give the results we  have obtained in $d=3$ within our approach
\cite{tissier01,delamotte03}. We  have computed the  value of $N_c(d)$
in this  dimension and found  $N_c(d=3)=5.1$.  Then, we  have searched
fixed points  {\it below}  this value  of $N_c$ in  order to  test the
results of  Pelissetto {\it et  al}  and the conjecture  of Calabrese
{\it et al}  that the physics of frustrated magnets would be governed
by fixed  points living  in the region  $N<N_c(d=3)$. This  search has
been unsuccessful.   In fact, within our approach,  we have discovered
that,  although there is  no fixed  point, there  exists a  {\it whole
region} in  the space of  coupling constants in  which the RG  flow is
slow. This generalizes the  notion of pseudo-fixed point introduced by
Zumbach  \cite{zumbach93}.  For  the  systems corresponding  to  these
coupling  constants, the  correlation lengths  are very  large  at the
transitions, that are  thus very weakly of first  order.  We have found
scaling  behaviors on  large ranges  of  temperature for  both XY  and
Heisenberg magnets  (see figure (\ref{log_m_xi})).   Moreover, we have
been able to account for the spreading of (pseudo-)critical exponents
that characterize these systems and, by varying the initial conditions
of the RG flow, to reproduce the critical exponents of the numerically
simulated  systems and  of many  compounds.  For  XY systems,  we have
found  $0.25<\beta<0.38$  and   $0.47<\nu<0.58$  and,  for  Heisenberg
systems,  $0.27<\beta<0.42$ and $0.56<\nu<0.71$,  ranges of  values in
which almost all experimental and  numerical values enter. Let us give
more  details.  We  first  consider  the XY  case.  According  to  the
experimental  values of  the critical  exponents, the  compounds split
into  two  groups:  STA   compounds  ---  CsMnBr$_3$,  CsNiCl$_3$  and
CsMnI$_3$  ---  for  which   one  has,  on  average,  $\beta=0.237(4),
\nu=0.555(21), \alpha=0.344(5)$ and $\gamma=1.075(42)$, and Helimagnets
compounds  ---   Holmium  and  Dysprosium   ---  for  which   one  has
$\beta=0.389(7), \nu=0.558(25)$ and  $\gamma=1.10(5)$ ($\alpha$ is not
given since it is poorly determined). Within our computations, we find
initial conditions of the flow  leading to critical exponents close to
those of  Helimagnets: $\beta=0.38$, $\nu=0.58$  and $\gamma=1.13$. We
also   easily  find  initial   conditions  leading   to  $\beta=0.25$,
corresponding  essentially  to  STA  compounds.  For  such  values  of
$\beta$, we  find that  $\nu$ varies between  0.47 and 0.49,  which is
somewhat below the  value found for CsMnBr$_3$. The  same splitting of
the  compounds  happens in  the  Heisenberg  case.  There is  a  group
composed        of        $\hbox{Cu(HCOO)}_2        2\hbox{CO(ND}_2)_2
2\hbox{D}_2\hbox{O},\hbox{Fe[S}_2                         \hbox{CN(C}_2
\hbox{H}_5)_2]_2\hbox{Cl},\hbox{VCl}_2$       and       $\hbox{VBr}_2$
characterized  by $\beta=0.230(8), \alpha=0.272(35),  \nu=0.62(5)$ and
$\gamma=1.105(21)$.    There    is    also    a    group    made    of
$\hbox{CsNiCl}_3,\hbox{CsMnI}_3$                                    and
$\hbox{CsMn(Br}_{0.19},\hbox{I}_{0.81})_3$  for which $\beta=0.287(9),
\alpha=0.243(3),  \nu=0.585(9)$  and  $\gamma=1.181(33)$.  Within  our
computations one can find  $\beta=0.27$ and $\nu=0.56$ which are
close to the experimental values of the second group. In contrast, we
have not  been able to  reproduce correctly the critical  exponents of
the first  group. This difficulty to reproduce  the critical exponents
originates in the different approximations we have used, in particular
the  truncations  in fields  and  derivatives.  Beyond this  technical
difficulty, the  trouble is that,  once the universality is  lost, the
pseudo-critical exponents  depend on the  specific compound considered
and their  precise determination would  require to know  precisely the
microscopic  structure  of  the   compounds  or  systems  studied  ---
providing the initial  conditions of the RG flow ---  and to take into
account  the full  field-dependence  of the  potential in  $\Gamma_k$.
Finally  and interestingly, we  have also  been able  to find,  in the
Heisenberg  and  in  the  XY  cases,  initial  conditions  leading  to
pseudo-critical exponents in good agreement with those obtained within
the  six-loop  calculation  \cite{pelissetto01a}.  In  the  Heisenberg
case, one can find exactly  the same values --- $\beta=0.30, \nu=0.55$
and  $\gamma=1.06$ ---  than Pelissetto  {\it et al} \cite{pelissetto01a}. 
 In  the XY  case, one
finds $\beta=0.33$, $\nu=0.56$ and  $\gamma=1.07$ which are very close
to   the   six-loop   results:  $\beta=0.31(2)$,   $\nu=0.57(3)$   and
$\gamma=1.10(4)$.

\begin{figure}[tbp] 
\centering                                        \makebox[\linewidth]{
\includegraphics[width=.5\linewidth,origin=tl]{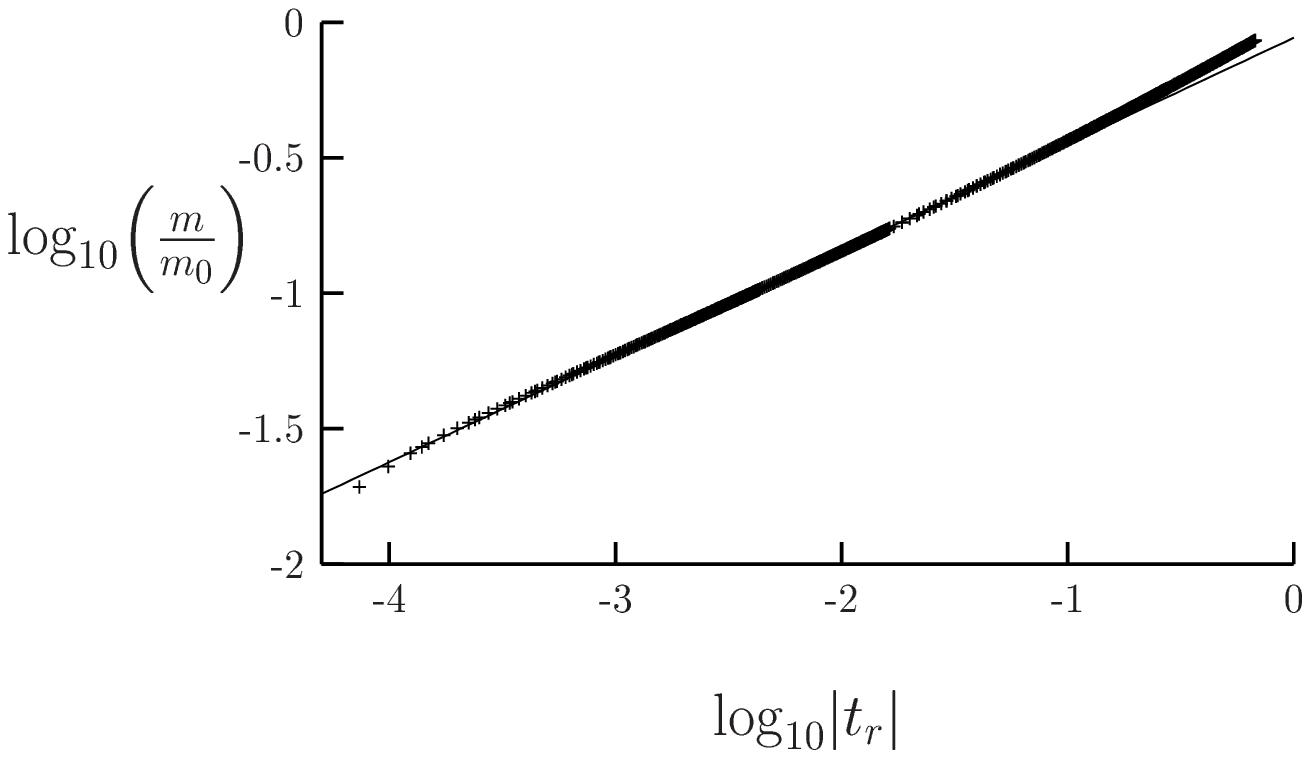}
\includegraphics[width=.5\linewidth,origin=tl]{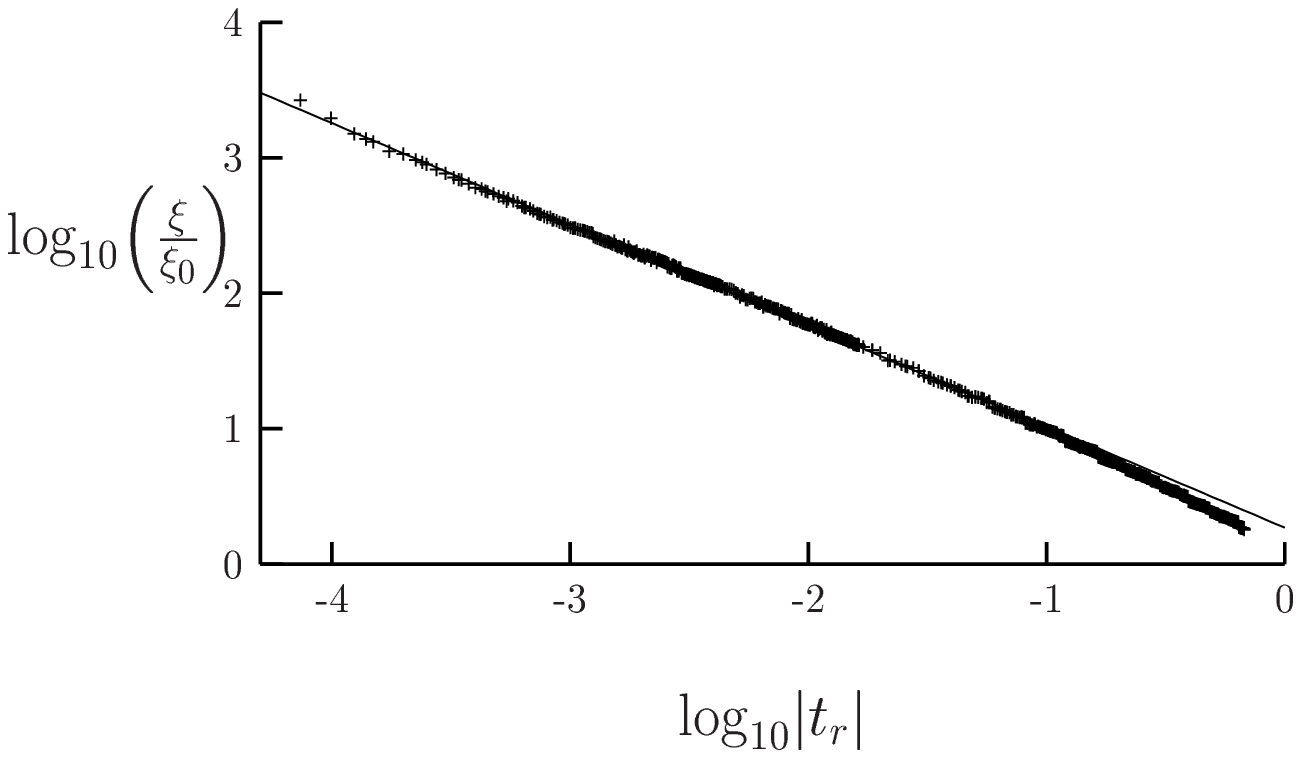}%
}
\caption{Log-log  plot   of  the  order  parameter  $m$   and  of  the
correlation  length  $\xi$  for  $N=3$  as functions  of  the  reduced
temperature $t_r$.}
\label{log_m_xi}
\end{figure}

\section{Conclusion}

On the basis  of their specific symmetry breaking  scheme, it has been
proposed  that the  critical physics  of frustrated  systems  in $d=3$
could be  characterized by critical  exponents associated with  a {\it
new}  universality  class. From  this  point  of  view, the  study  of
frustrated magnets has been rather disappointing, the experimental and
numerical contexts excluding such an hypothesis. At the same time, the
phenomenology  of frustrated  magnets has  displayed a  novel  kind of
critical  behavior:  the  existence  of  {\it  generic}  scaling  {\it
without}    universality~\cite{tissier01,delamotte03}.    Within   the
framework   of  our   nonperturbative  approach,   this   generic  and
nonuniversal scaling  finds a natural explanation in  terms of slowness
of the flow.  This method has also explained  the mismatch between the
different  perturbative   approaches  by  means  of   a  mechanism  of
annihilation  of  fixed points  that  invalidates the  low-temperature
perturbative approach  performed on the NL$\sigma$ model.  It would be
satisfying  to  understand  the  very  origin  of  this  failure.  The
influence  of  nontrivial  topological  configurations  on  the  phase
transition in  $d=3$ has been  invoked. It remains however  to confirm
that  these configurations  indeed play  such  a crucial  role and  to
understand  whether  they  are  really  responsible  for  the  first  order
character of the transitions. Finally,  and even though  it could appear as
a rather formal  question, it would be also  satisfying to clarify the
discrepancy between the  nonperturbative and the six-loop perturbative
approaches.  The  matching between the sets of  exponents found within
these different approaches suggests  that there exists a common origin
to these  two sets of exponents.   One can easily  imagine that, while
the two  kinds of  computations would be  almost converged as for the
determination of  the exponents themselves, one  of these computations
would not be converged as for  the true nature --- real or complex ---
of the  fixed points.  This question will  be likely clarified  in the
near future.

{\it Note added in proof.}  Let us mention  that a Monte Carlo study \cite{peles2}
has confirmed the existence of first order  behaviors  for  a whole
family of XY frustrated magnets, in agreement with our scenario.

\end{document}